\documentclass[preprint,12pt]{aastex}

\shorttitle{Time-Distance Far-Side Imaging} 
\shortauthors{Zhao}

\begin{document}
\title{Time-Distance Imaging of Solar Far-Side Active Regions}
\author{Junwei Zhao} 
\affil{W.~W.~Hansen Experimental Physics Laboratory, Stanford University,
Stanford, CA94305-4085}

\begin{abstract}
It is of great importance to monitor large solar active regions 
in the far-side of the Sun for space weather forecast, in particular, 
to predict their appearance before they rotate into our view from 
the solar east limb. Local helioseismology techniques, including 
helioseismic holography and time-distance, have successfully imaged 
solar far-side active regions. In this Letter, we further explore 
the possibility of imaging and improving the image quality of 
solar far-side active regions by use of time-distance helioseismology. 
In addition to the previously used scheme with four acoustic signal 
skips, a five-skip scheme is also included in this newly developed 
technique. The combination of both four- and five-skip far-side 
images significantly enhances the signal-to-noise ratio in the far-side
images, and reduces spurious signals. The accuracy of the far-side 
active region imaging is also assessed using one whole year's solar 
observation.
\end{abstract}


\section{INTRODUCTION}
Being able to detect a large solar active region before it rotates into 
our view from the Sun's far-side, and monitor a large active region 
after it rotates out of our view into the far-side from the Sun's 
west limb, is of great importance for the space weather forecast 
\citep{sch03}. Using helioseismic holography technique \citep[for 
a review, see][]{lin00a}, \citet{lin00b} mapped the central 
region of the far-side Sun by utilizing double-skip solar acoustic 
signals visible in the near-side of the Sun but originated from 
or arriving into the targeted far-side area. Then, \citet{bra01} 
developed furthermore this technique to map the near-limb and polar 
areas of the solar far-side by combining acoustic signals visible
in the near-side after single and triple skips on either side of 
the targeted far-side area. These efforts have made monitoring
active regions of the whole back side of the Sun possible, and daily
solar far-side images have become available using {\it Solar 
and Heliospheric Observatory}/Michelson Doppler Imager 
\citep[{\it SOHO}/MDI;][]{sch95} and Global Oscillation Network Group 
\citep[GONG;][]{har96} observations\footnote{See
http://soi.stanford.edu/data/farside/ for MDI results, and see
http://gong.nso.edu/data/farside/ for GONG results}. That the far-side 
active regions are detectable by phase-sensitive holography \citep{bra00} is
mainly because low- and medium-$l$ ($l$ is spherical harmonic degree) 
acoustic waves manifest apparent travel time deficit in solar active 
regions, and it is these travel time anomalies that far-side maps 
image, although it is still quite uncertain whether such anomalies 
are due to the near-surface perturbation in solar magnetic regions
\citep{fan95, lin05}, or caused by faster sound speed structures in the 
interior of those active regions \citep[e.g.,][]{kos00, zha06}. 

Time-distance helioseismology is another local helioseismological tool 
that is capable of mapping the solar far-side active regions. It was
demonstrated that time-distance helioseismology could map the central
areas of the far-side Sun in a similar way as \citet{lin00b} had done
\citep{duv00, duv01}, however, no studies have been carried out to 
extend the mapping area into the whole far-side. In this Letter, 
utilizing the time-distance helioseismology technique, I explore 
both the previously used four-skip scheme and a new five-skip scheme 
and image two whole far-side maps using four-skip and five-skip
acoustic signals separately. Combining both maps gives us solar 
far-side images with better signal-to-noise ratio and more reliable 
maps of active regions on the far-side. This provides another solar 
far-side imaging tool in addition to the existing holography technique, 
hence gives a possibility to cross-check the active regions seen in 
both techniques, and enhances the accuracy of monitoring solar 
far-side active regions.

\section{DATA AND TECHNIQUE} 
The medium-$l$ program of {\it SOHO}/MDI provides nearly continuous 
observations of solar oscillation modes with $l$ from $0$ to $\sim$300. 
The medium-$l$ data are acquired with one minute cadence and a 
spatial sampling of 10 arcsec (0.6 heliographic degrees per pixel) 
after some averaging onboard the spacecraft and ground pipeline processing 
\citep{kos97}. The oscillation modes that this program observes and the 
continuity of such observations make the medium-$l$ data ideal to be 
used in solar far-side imaging. 

\begin{figure}
\plotone{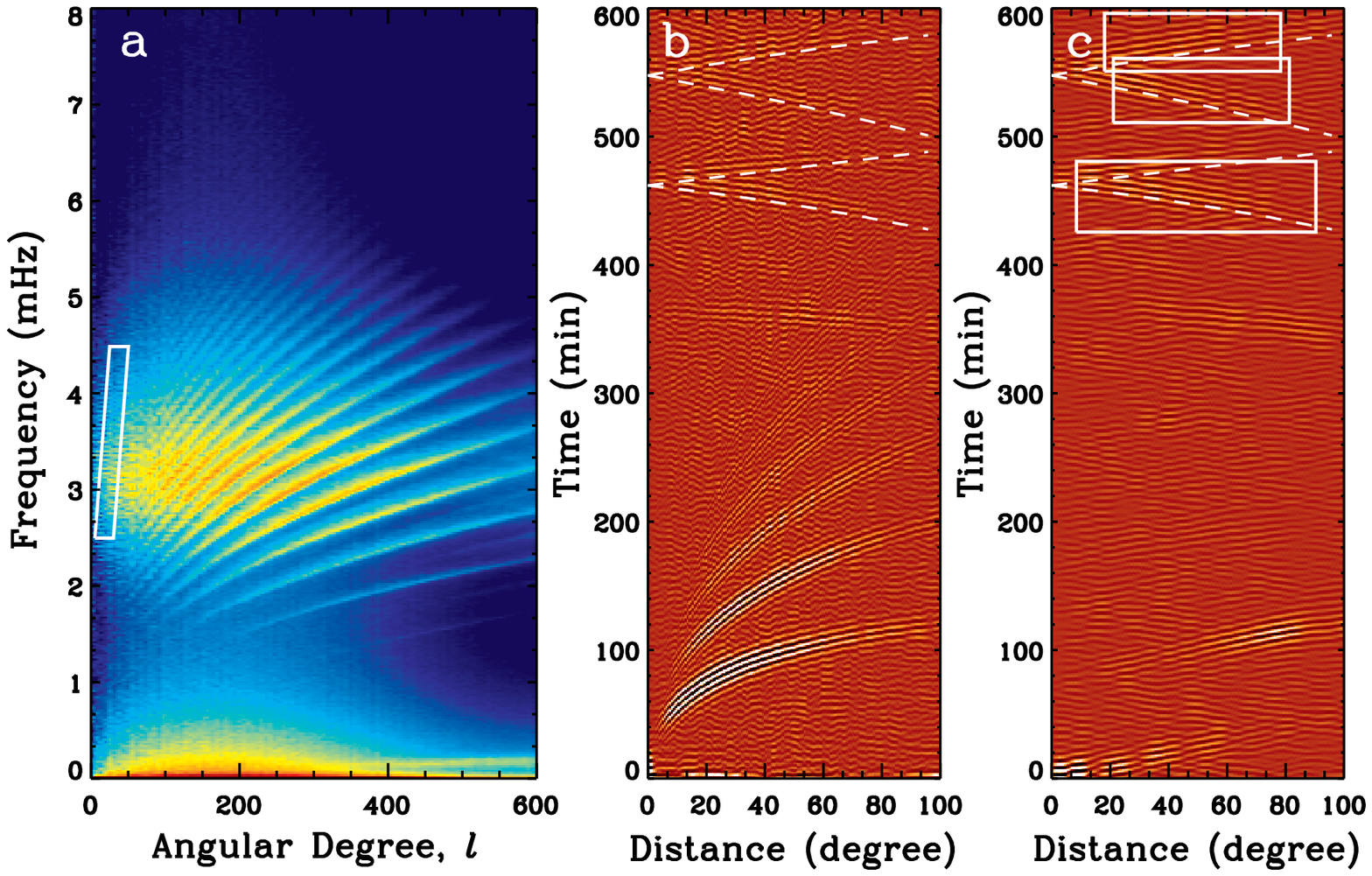}
\caption{Power spectrum diagram ($a$) computed from a 1024-min MDI 
medium-$l$ data set, time-distance diagram ($b$) computed using the whole 
power spectrum of the same data set, and time-distance diagram ($c$) 
computed using only the oscillations that have frequency and $l$ 
included in the white quadrangle as indicated in ($a$). The white 
dashed lines in both ($b$) and ($c$) are theoretical time-distance 
relationships based on acoustic ray approximation, with the lower 
`$<$'-like curve as the fourth-skip, and the upper `$<$'-like 
curve as the fifth-skip. The lower and upper branch of each 
`$<$'-like curve represents acoustic wave propagation distance 
shorter and longer than $360\degr$, respectively. The white boxes 
in ($c$) delimit the acoustic travel distances and times used for 
far-side imaging.} 
\label{fg1}
\end{figure}

The acoustic wave signals that originate from and then return to 
the solar front side after traveling to the back side and 
experiencing four and five bounces are to be used in the 
time-distance far-side analysis. It is natural and often useful 
to first examine whether such acoustic signals are detectable by 
the time-distance technique. A 1024-min MDI medium-$l$ data set 
with a spatial size of $120\degr\times120\degr$ is used for such 
an examination after the region is tracked with Carrington rotation 
rate and is remapped using Postel's projection with the solar disk 
center as its remapping center. The $l-\nu$ power spectrum
diagram of this data set is shown in Figure~\ref{fg1}$a$. The 
time-distance diagram computed using all oscillation modes 
(Figure~\ref{fg1}$b$) clearly shows acoustic bounces of 
up to seven times in the front side, and at the locations of the 
expected signals traveling back to the front side 
after four and five bounces, there are some not-very-clear but existing 
signals. Since the acoustic signals that are needed for the far-side 
analysis should have a fairly long single-skip travel distance, thus 
corresponding to low-$l$ modes, it is helpful to filter out all other 
acoustic modes and keep only those signals that are corresponding to 
the required travel distances. Such a filtering should help to improve
the signal-to-noise ratio of time-distance signals at the desired 
distances. However, such a filtering is applied to the data 
after Postel's projection, thus it only works best where the 
geometry is least distorted. The white quadrangle in Figure~\ref{fg1}$a$ 
delimits the acoustic modes that are used in both four- and five-skip 
analyses (note that when computing far-side images, both four- and 
five-skip analyses use only part of the acoustic modes shown in the 
white quadrangle), covering a frequency range of $2.5 - 4.5$ mHz and 
$l$ of $3 - 50$, and the time-distance diagram made using only these 
modes (Figure~\ref{fg1}$c$) shows clearly the four- and five-skip 
time-distance signals. Considering that the time-distance group travel
time is a bit off from the ray approximated time, as was pointed out 
by \citet{kos96} and \citet{duv97}, it is understandable that theoretical
travel time is a few minutes off from the time-distance group travel
time after four and five bounces.

In order to keep good signal-to-noise ratio as well as reasonable
computational burden, not all acoustic signals that come
back from the far-side after four or five skips are used. The white
boxes in Figure~\ref{fg1}$c$ indicate the distances that are utilized 
for far-side imaging. Namely, as sketched in Figure~\ref{fg2}, for 
the four-skip scheme, the time-distance annulus radii are $133\fdg8 - 
170\fdg0$ from the target point for the double-double skip combination 
that is used to map the far-side central area, and $66\fdg9 - 85\fdg0$ 
for the single-skip and $200\fdg7 - 255\fdg0$ for the triple-skip in 
the single-triple combination that is used to map the areas that are close 
to the far-side limb and polar regions; such a combination covers a 
total of $190\degr$ in longitude (at the equator, $5\degr$ past the limb 
to the front side near both limbs). For the five-skip scheme, the annulus 
radii are $111\fdg6 - 174\fdg0$ from the target point for 
the double-skip and $167\fdg4 - 261\fdg0$ for the triple-skip; this
scheme covers a total of $160\degr$ in longitude, less than the 
whole far-side. Unlike in \citet{bra01} where far-side polar areas 
were also included in the imaging, to reduce unnecessary computations
only areas lower than the latitude of $48\degr$, where nearly all 
active regions are located, are included in the time-distance far-side 
imaging computations reported here.

The computation procedure is as follows. A 2048-min long MDI medium-$l$ 
data set is tracked with Carrington rotation rate, and remapped 
to Postel's coordinates centered on the solar disk as observed at
the mid-point of the dataset, with a spatial sampling of $0\fdg6$ 
pixel$^{-1}$, covering a span of $120\degr$ along the equator as
well as the central meridian. This data set is 
then filtered in the Fourier domain to keep only the oscillation 
modes that have travel distances in agreement with the distances 
listed above. Corresponding pixels in the annuli on both sides 
of the target point are selected, and the cross-covariances with 
both positive and negative travel time lags are computed. Such 
computations for all distances shown in the white boxes of 
Figure~\ref{fg1}$c$ are repeated. Then, both time lags are 
combined and all cross-covariances obtained from different 
distances after appropriate shifts in time that are obtained from 
theoretical estimates 
are also combined. The final cross-covariance is fitted with a 
Gabor wavelet function \citep{kos96}, and an acoustic phase travel 
time is obtained. This procedure is repeated for the double-double 
and single-triple combinations in the four-skip scheme, and for 
the double-triple combination for total distances both shorter and longer 
than $360\degr$ in the five-skip scheme, separately. The far-side image is
a display of the measured acoustic travel times after a Gaussian
smoothing with FWHM of $2\fdg0$.

\section{RESULTS}
NOAA AR0484, AR0486 and AR0488 are selected to test this time-distance
far-side imaging technique. Figure~\ref{fg3} shows the MDI synoptic 
magnetic field chart when these active regions were on the near-side 
of the Sun before they rotated into the far-side. AR0488
was still growing when the magnetic field data to make this synoptic
map were taken. 

Selected far-side images of these three active regions, obtained from 
separate four- and five-skip measurement schemes, and a combination 
of both measurements, are shown in the top three rows of Figure~\ref{fg4}. 
In each image, the near-side magnetic field map is combined with 
the far-side acoustic travel time map, and the combined map is 
displayed based on Carrington longitudes. Standard deviations, $\sigma$,
are 4.1 sec and 4.5 sec for four-skip and five-skip far-side maps,
respectively, and after combination of both measurement schemes, 
$\sigma$ falls to 3.3 sec. In these maps, the three interested active
regions are mostly visible in all four-skip, five-skip and combined 
results, except when the regions fall into the uncovered areas of 
five-skip measurement. 

The bottom row of Figure~\ref{fg4} presents the same map as the third
row but highlighting active regions by Gaussian smoothing the unsigned
near-side magnetic field, and displaying the far-side travel time map with 
a threshold of $-3.5\sigma$ to $-2.0\sigma$, i.e., $-11.5$ to $-6.5$ sec.
It is clear that the images combining four- and five-skip results 
are quite clean of spurious signals, although unidentified 
features still exist. It is also noteworthy that the far-side 
images clearly show these active regions when part of the regions were in 
the far-side and part in the near-side, like AR0486 and AR0488 in 
the fourth row and third column of Figure~\ref{fg4}.

Both four- and five-skip far-side acoustic travel times are displayed 
after a mean travel time background is removed. It turns out that 
the background mean travel time depends on its angular distance
from the antipode of the solar disk center. Figure~\ref{fg5} presents
the variation of background mean travel times at different locations for 
both measurement schemes after the mean value of the background 
is removed. Basically, 
the mean acoustic travel times vary with a magnitude of $\sim$4 sec 
for four-skip measurement, and a magnitude of $\sim$9 sec for five-skip
measurement. The measured mean acoustic travel times are shorter for 
the four-skip scheme, but longer for the five-skip scheme when the 
targeted areas are near the limb and near the antipode of the solar
disk center. These variations are unlikely physical, yet it is unclear
why there are such variations in the measurement. The rms of travel
times can be used to assess the quality of far-side images. Also shown
in Figure~\ref{fg5} is travel time rms variation with distance to
the far-side center. It shows that double-double scheme 
has the lowest rms near the far-side center, while the single-triple 
scheme has the lowest rms near the far-side limb. For the five-skip
measurement, it has the lowest rms when at and roughly $30\degr$ from
the far-side center, and the rms increases substantially close to the limb.

It is interesting to make a statistical study investigating how 
accurate the far-side active region imaging is. Since sometimes 
active regions change fast, grow or decay unexpectedly, it 
is quite impossible to make a precise assessment on the accuracy of 
far-side imaging. Therefore, the results given below can only be 
regarded as a reference.  The far-side images of the whole year of 
2001 were computed at 12-hour intervals, and a total of 730 images 
were obtained and displayed in the same way as the bottom row of 
Figure~\ref{fg4}. Typically, a solar area is located on the far-side 
of the Sun for about 13.5 days, or 27 far-side images. Far-side 
images are often noisy, and even for a large active region, 
it is quite unlikely for it to be unambiguously visible in each of 
27 images. Also, active regions on the far-side grow and decay as well. 
For these reasons, if an area is visible as dark in our figures 6 times 
(or on 3 days) on the other side, where 6 is an arbitrary number, that 
area is regarded as an active region. Based on this assumption, it is found
that for 59 active regions that rotate into the near-side from the east limb,
53 (or $89.8\%$) are able to be detected far-side, and for 63 active
regions that rotate into the far-side from the west limb, 56 (or 
$89.0\%$) are detectable on the far-side. Among these regions, 
33 are actually detected both before and after the near-side appearance. 
And, from the other perspective, for 61 regions that are detected by 
far-side imaging rotating into the far-side from the west limb, 55 (or
$90.1\%$) are visible near the west limb of the near-side; for 
53 active regions that are detected by far-side imaging rotating out 
of the far-side into the near-side, 49 (or $92.4\%$) are visible near 
the east limb of the near-side. Among these regions, 28 are visible on 
the near-side before and after the region is located in the far-side. 
These statistics do not count all active regions that appeared in that 
year, but only those that have a diameter roughly larger than $8\fdg0$ 
after projected into a spatial resolution of solar disk center and 
viewed in a figure like the fourth row of Figure~\ref{fg4}, when these 
regions are near either limb of of the Sun.

\section{DISCUSSION}
By combining four-skip and five-skip measurement schemes, I have 
successfully made time-distance far-side 
images of the Sun with a fast computation speed. The combination 
significantly enhances the signal-to-noise ratio of far-side acoustic 
travel times over either four- or five-skip measurement, thus making 
the composite far-side active region map much cleaner. It helps to remove 
most, but not all of the spurious features that signify active 
regions. On the other hand, the combination of both maps 
also remove some small active regions that can otherwise be seen 
in one map or the other.

Time-distance far-side imaging provides another whole far-side imaging 
tool in addition to the existing helioseismic holography technique. 
It gives results that are in reasonable agreement with the holography 
results, which can be seen at {\it SOHO}/MDI website, by visual 
comparison, yet detailed comparisons are not done. It is intriguing 
to calibrate the measured far-side acoustic travel times into 
magnetic field strength, and some efforts have been taken using 
holography results \citep{gon07}. However, there are some apparent 
difficulties to carry out the calibration, because the far-side images of 
active regions are often relatively noisy, and sizes and shapes 
of those regions change from image to image. In addition, the magnetic 
field strengths of the far-side active regions are also unknown, 
although the full sphere magnetic field made by use of flux 
dispersal model \citep{sch03} may help in this. Perhaps eventually, 
a numerical simulation of solar oscillations in the whole Sun is 
needed to determine a satisfactory calibration.

\begin{figure}
\plotone{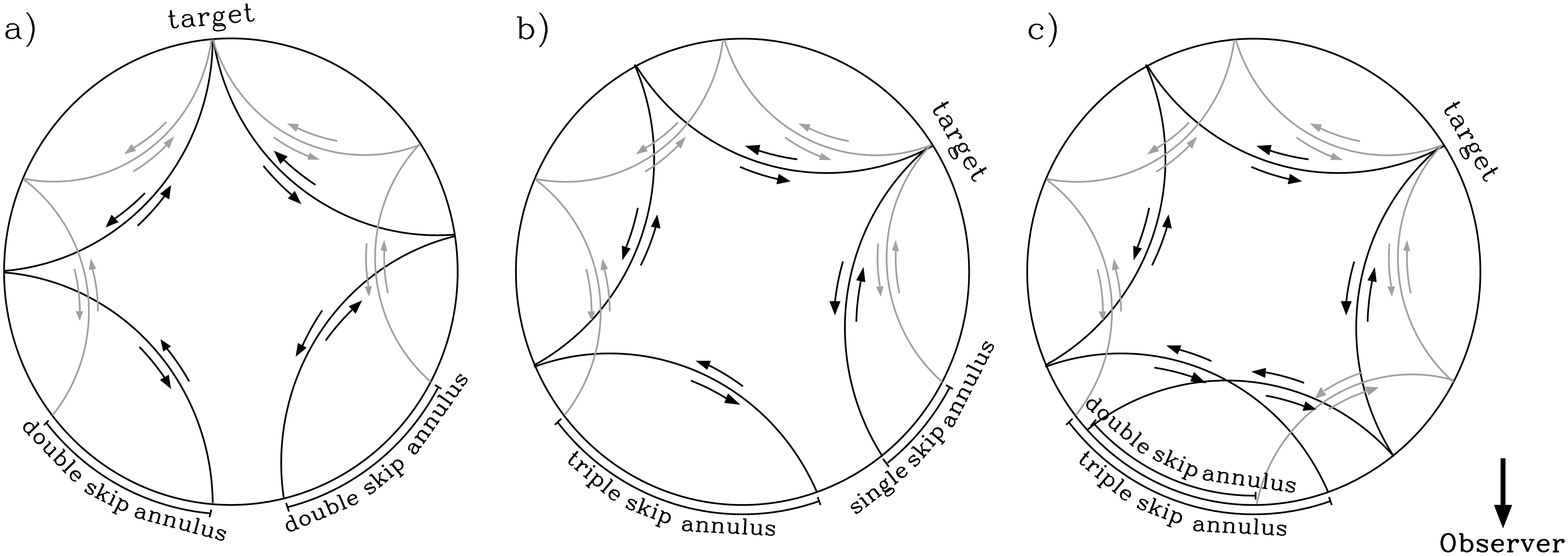}
\caption{Sketches for four-skip measurement schemes, which includes 
($a$) double-double skip combination when the target point is near 
the central area of the far-side and two sets of double-skip rays 
are located on both sides of the target point, and ($b$) single-triple
skip combination when the target point is near the limb or polar area
of the far-side and one set of single-skip and one set of triple-skip
rays are located on either side of the target; and, five-skip measurement
schemes ($c$) when one set of double-skip and one set of triple-skip 
rays are located on the either side of the target point. }  
\label{fg2}
\end{figure}

\begin{figure}
\epsscale{0.7}
\plotone{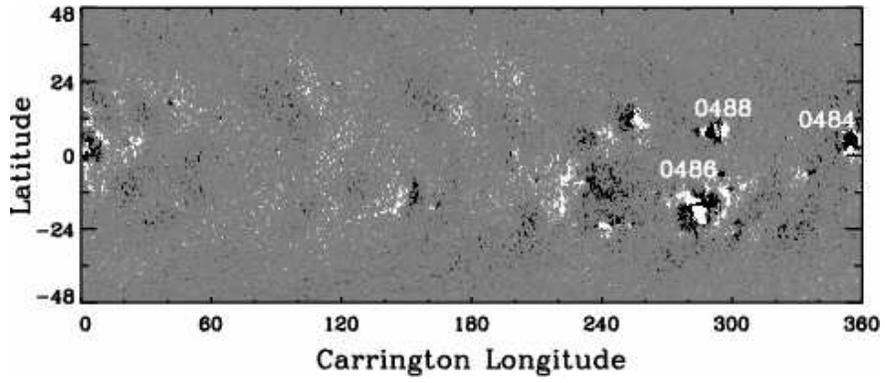}
\caption{MDI magnetic field synoptic chart for Carrington rotation 2009, 
taken from October 23 to November 19, 2003, and made by use of 
observations near the central meridian only. Note that AR0484 is 
divided at the longitude of $0\degr$ and $360\degr$. }
\label{fg3}
\end{figure}

\begin{figure}
\epsscale{1.0}
\plotone{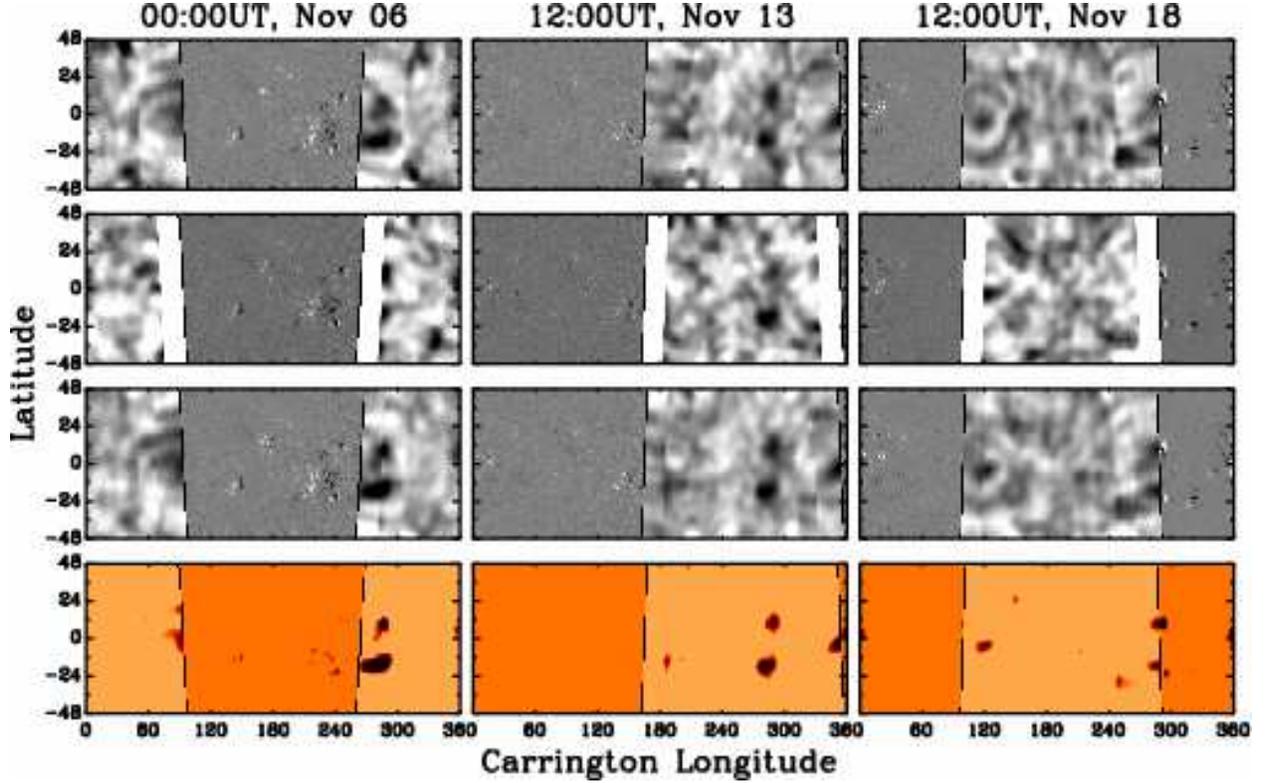}
\caption{Results of time-distance far-side active region imaging, obtained
from four-skip ({\it first row}), five-skip ({\it second row}), and 
combination of four- and five-skip measurements ({\it third and bottom
row}). From the left to the right column, images were obtained at 00:00UT of 
November 6, 12:00UT of November 13, and 12:00UT of November 18, 2003, 
respectively. The given observation time for far-side image is the 
middle time of its 2048-min observational period. Each black and white 
image displays a combination of the near-side MDI magnetic field map 
and a time-distance far-side acoustic travel time map. The boundaries
between the far-side and near-side are not vertical because of the solar
B-angle adjustment. Magnetic field is displayed with a range of $-150$
to 150 Gs, and the acoustic travel time ranging from $-12$ to 12 sec. 
Blank regions in five-skip images indicate areas that cannot be covered 
by this measurement. Color images display the near-side map with
a range of 40 to 150 Gs after a Gaussian smoothing of unsigned magnetic 
field with a same FWHM that is applied to travel time maps, and 
display the far-side map with a range of $-3.5\sigma$ to $-2.0\sigma$ 
in order to highlight the far-side active regions that are of our 
interest. To better distinguish far-side and near-side images, different 
color contrast is used. }
\label{fg4}
\end{figure}

\begin{figure}
\epsscale{0.7}
\plotone{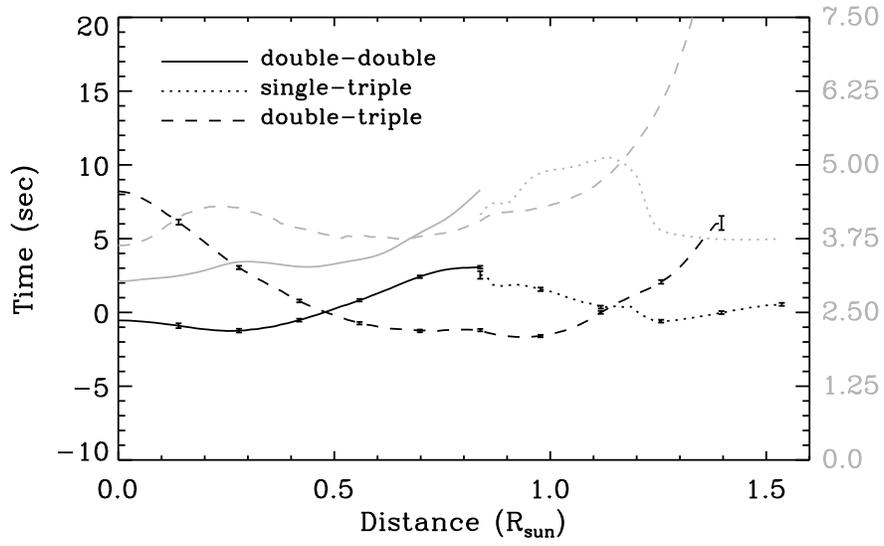}
\caption{Background travel time variations ({\it black curves}) and travel
time rms variations ({\it grey curves}) as functions of angular distance
from the antipode of solar disk center for different combinations of 
acoustic wave bounces. Scales for grey curves are marked on the right
hand side of the figure.} 
\label{fg5}
\end{figure}

It is notable that the five-skip measurement scheme has a large fraction
of overlapping areas as the four-skip measurement in the annulus of one side,
and has one more skip on the other side (see Figure~\ref{fg2}), but 
results from both measurement schemes have quite uncorrelated noises
(see Figure~\ref{fg4}). It is unlikely that the noise differences of 
two schemes are caused while waves travel merely one more skip, $1/5$
of the total distance, however, it may imply noises come from data 
or the filtering of geometrically distorted data.

The holography far-side images show strong acoustic travel time variation
from the far-side center to limb, with an order of 70 sec or so in 
their four-skip measurement scheme (P. Scherrer, private communication),
and this is believed (C. Lindsey, private communication) to be connected 
with the ``ghost signature'' described by \citet{lin04}. Although such a 
travel time variation trend also exists in time-distance far-side 
images, the variation magnitude is merely 4 sec in four-skip scheme, 
substantially smaller than the holography trend. The variation in the 
five-skip scheme measurement is about twice of that in the four-skip 
scheme. What causes such travel time variations in our measurement, 
though small, is worth further investigation.

With the availability of both time-distance and helioseismic holography
far-side images, it is more robust to monitor activities on the 
far-side of the Sun, and we can forecast the appearance of large 
active regions rotating into our view from the far-side with more 
confidence. Furthermore, the success in far-side imaging provides
us some experiences and confidence in analyzing low- and medium-$l$ 
mode oscillations by use of local helioseismology techniques, and 
this will greatly help us in analyzing solar deeper interiors and 
polar areas by use of these techniques.

\acknowledgments
I thank Tom Duvall and Sasha Kosovichev for reading through the 
manuscript and giving valuable comments to improve this paper,
as well as their suggestions while developing the code. I am deeply
indebted to Charlie Lindsey and the referee, Doug Braun, for 
thoroughly studying my manuscript and giving numerous constructive 
comments. I also thank Phil Scherrer for encouraging me to 
carry out this work. {\it SOHO} is a project of international 
cooperation between ESA and NASA.

\clearpage

\clearpage

\clearpage

\clearpage

\clearpage

\end{document}